\documentclass[aps,prd,11pt]{revtex4}
\usepackage[colorlinks=true, pdfstartview=FitV, linkcolor=blue, citecolor=red, urlcolor=magenta]{hyperref}

\usepackage{graphicx}
\usepackage{latexsym}
\usepackage{amsmath}
\usepackage{amsfonts}
\usepackage{amssymb}
\usepackage{verbatim}

\newcommand{\be}{\begin{equation}}
\newcommand{\ee}{\end{equation}}
\newcommand{\ben}{\begin{eqnarray}}
\newcommand{\een}{\end{eqnarray}}

\begin{document}

\title{Brane structure from a scalar field in general covariant Horava-Lifshitz gravity}
\author{D. Bazeia$^{a}$, F.A. Brito,$^{b}$ and F.G. Costa,$^{a,c}$ } 
\affiliation{
$^{a}$Departamento de F\'\i sica,
Universidade Federal da Para\'\i ba, Caixa Postal 5008,
58051-970 Jo\~ ao Pessoa, Para\'\i ba,
Brazil\\
$^{b}$Departamento de F\'\i sica,
Universidade Federal de Campina Grande, Caixa Postal 10071,
58109-970  Campina Grande, Para\'\i ba,
Brazil\\
$^{c}$Instituto Federal de Educa\c c\~ao, Ci\^encia e Tecnologia da Para\'\i ba (IFPB), Campus  Picu\'i, Brazil\\
}

\begin{abstract}
In this paper we have considered the structure of the non-projectable Horava-Melby-Thompson (HMT) gravity to find braneworld scenarios. A relativistic scalar field is considered in the matter sector and we have shown how to reduce the equations of motion
to first-order differential equations. In particular, we have studied thick brane solutions of both the dilatonic and Randall-Sundrum types.
  
\end{abstract}\maketitle

\section{Introduction}

The brane scenario in higher-dimensional theories is being investigated as a candidate for solving some fundamental problems in high energy physics such as hierarchy, cosmological constant and others. The formulation of the Randall-Sundrum model \cite{rand}, is based in terms of a single infinite extra dimension and the physical world appears as a four-dimensional spacetime embedded into an anti-de Sitter ($AdS$) space. In this scenario we can add scalar fields with usual dynamics which are also supposed to depend only on the extra dimension and allow them to interact with gravity in the standard way \cite{gold-wise}. The study of scalar fields coupled to gravity in warped geometries has been frequently reported in the literature \cite{cs,varios1,varios2,varios3,varios4}. On the other hand, Horava has proposed a new theory of gravitation \cite{horava} which has been extensively studied (See \cite{blas2} for a review). This theory, commonly referred to as Horava-Lifshitz (HL) gravity, is a nonrelativistic theory with an anisotropic scaling symmetry of space and time. The addition of higher spatial derivative terms in the action without their time derivative counterparts renders the theory power counting renormalizable. In order to restore the diffeomorphism symmetry at low energies, the theory is supposed to flow dynamically from a scale invariant theory in the ultraviolet (UV) to General Relativity (GR) in the infrared (IR) limit. The theory exhibits an anisotropic scaling between space and time given by
\be x\rightarrow \ell x,\;\;t\rightarrow\ell^zt,\ee
where $z=4$ in the $(4+1)$-dimensional spacetime for power-counting renormalizability \cite{horava}.

The gauge symmetry of the theory is broken down to the foliation-preserving diffeomorphism, Diff(M,F),
\be \delta t=-f(t),\;\; \delta x^{i}=-\zeta^{i}(t,x).\ee
The dynamical variables are the lapse function ($N$), the shift function ($N_{i}$) and the spatial metric $g_{ij}$ (roman letters indicate spatial
indices). In terms of these fields the full metric is written as an ADM decomposition as follows
\be ds^{2}=N^{2}dt^{2}-g_{ij}(dx^{i} + N^{i}dt)(dx^{j} + Njdt).\ee
The variables $N$, $N_{i}$ and $g_{ij}$ transform as 
\be\nonumber \delta N = \zeta^{k}\nabla_{k}N+\dot{N}f+N\dot{f},\ee
\be\nonumber \delta N_{i}=N_{k}\nabla_{i}\zeta^{k}+\zeta^{k}\nabla_{k}N_{i}+g_{ik}\dot{\zeta}^{k}+\dot{N}_{i}f+N_{i}\dot{f},\ee
\be \delta g_{ij}=\nabla_{i}\zeta_{j}+\nabla_{j}\zeta_{i}+f\dot{g}_{ij}.\ee

In these equations $\dot{f}\equiv df/dt$ and $\nabla_{i}$ denotes the covariant derivative with respect to $g_{ij}$. From these expressions one can see that $N$ and $N_{i}$ play the role of gauge fields of the Diff(M, F). Thus, it is natural to assume that $N$ and $N_{i}$ receive the same dependence on space and time as the corresponding generators \cite{horava}, 
\be N=N(t),\;\; N_{i}=N_{i}(t, x),\ee
which is often referred to as the projectability condition.

The Diff(M, F) diffeomorphisms lead to one more degree of freedom (a spin-$0$ graviton) in the gravitational sector that needs to be decoupled from the IR regime to be consistent with observations \cite{blas2, b.pere, s.muk}. Considerations in cosmology were given in \cite{cosmovarios}. An interesting approach is to eliminate the spin-$0$ graviton by introducing the U(1) gauge field $A$ and the Newtonian `prepotential', by extending the Diff(M, F) symmetry to include a local $U(1)$ symmetry \cite{hmt}. Another approach is to abandon the projectability condition. In the `non-projectable theory' the lapse function is allowed to depend on space $N = N(x^{i}, t)$ and one may include, in the action, the vector field \cite{blas}
\be a_{i} = \nabla_{i}\ln(N).\ee
The presence of this vector field solves the instability and strong coupling problems, however leads to a proliferation of independent coupling constants \cite{kimpton}. According to \cite{m.li}, the violation of the projectability condition often leads to the inconsistency problem, though this is not the case in the setup of \cite{blas} --- see \cite{j.kluson} for further related discussions.

An extended version of HL gravity without the projectibility condition but with the enlarged symmetry was proposed in \cite{t.zhu}. So one can reduce significantly the number of the independent coupling constants presented in the version of the non-projectable HL theory. On the other hand, it was also allowed a softly breaking in the detailed balance condition. This procedure turns the theory to be both UV complete and IR healthy. By implementing the enlarged symmetry one can eliminate the spin-$0$ graviton and all the problems related, such as the instability and strong coupling in the pure gravity sector.

Some aspects of the HL theory in $5$ dimensions was explored in the literature. In particular, in reference \cite{benfica} was investigated a braneworld scenario in a Horava-like five-dimensional theory at warped spacetimes. In this paper, we study brane structure with a single scalar field in the Horava and Melby-Thompson (HMT) setup with the \textit{non-projectability} condition. 

The paper is organized as follows. In Sec.~\ref{Sec1}, we shall give a brief introduction on  HMT setup with the non-projectability condition. In Sec.~\ref{Sec2} 
we introduce the setup for studying braneworld scenarios. In Sec.~\ref{Sec3} we shall consider a relativistic scalar field in the matter sector. We find explicit braneworld solutions. 
Finally in Sec.~\ref{conclu} we make our final considerations.

\section{Horava-like model in five dimensions without projectability condition}%%%%%%%%%%%%%%%%%%%%%%%%%%%%%%%%%%%%%%%%%%%%%%%%%%%%%%%%%%%%%%%%%
\label{Sec1}

In this section, let us give a brief introduction to the HMT setup \cite{hmt} with the non-projectability condition \cite{blas,m.li,j.kluson}. The full action of the theory is given by
\be\label{theroy1} S=\zeta^2\int dtdx^{3}dwN\sqrt{g}\left(L_{K}-L_{V}+L_{\varphi}+L_{A}+L_{\lambda}+\frac{1}{\zeta^2}L_{m}\right).\ee
We shall define the five-dimensional vector $X^{\Omega}=(t,x^{I},w)$, where, $\Omega=0,1,2,3,4$ denote $5D$ spacetime indices,  with $x^{0}=t$, $I=1,2,3$. The spatial part is denoted by $x^{i}$ with $i=1,2,3,4$ and $x^{4}=w$. In addition, we are considering $g=\det{(g_{ij})}$.

The kinetic term, $L_{K}$, in the action is given by
\be\label{Kinetic} L_{K}=K_{ij}K^{ij}-\lambda K^2.\ee
Furthermore, we have
\be\nonumber L_{\varphi}=\varphi\Theta^{ij}(2K_{ij}+\nabla_{i}\nabla_{j}\varphi+a_{i}\nabla_{j}\varphi)\ee
\be\nonumber+(1-\lambda)[(\nabla^2\varphi+a_{i}\nabla^{i}\varphi)^2+2(\nabla^2\varphi+a_{i}\nabla^{i}\varphi)K]\ee
\be\nonumber +\frac{1}{3}\hat\Theta^{ijkl}[4(\nabla_{i}\nabla_{j}\varphi)a_{(k}\nabla_{l)}\varphi+5(a_{(i}\nabla_{j)}\varphi)a_{(k}\nabla_{l)}\varphi\ee
\be +2(\nabla_{(i}\varphi)a_{j)(k}\nabla_{l)}\varphi+6K_{ij}a_{(l}\nabla_{k)}\varphi].\ee
In order to eliminate the spin-0 gravitons one needs to consider $U(1)$ gauge invariance in the general
action of the gravitational part of the HL gravity \cite{hmt} --- see also \cite{apendix}. Thus, by adding the term 
\be L_{A}=\frac{A}{N}(2\Lambda_{g}-R),\ee
one can introduce a $U(1)$ gauge field $A$, which transforms as 
\be
\delta_\alpha A=\dot{\alpha}-N^i\nabla_i\alpha
\ee
accompanied with the gauge transformation of the Newtonian prepotential $\varphi$
 \be
 \delta_\alpha\varphi=-\alpha,
 \ee
in order for the theory to have the $U(1)$ symmetry, 
where $\alpha$ denotes the $U(1)$ generator. Since the action is invariant under such transformation we shall fix the gauge, for later convenience, at $A=0$ and $\varphi=0$ in the equations of motion. For arbitrary $\lambda$ we can write
\be L_{\lambda}=(1-\lambda)[(\Delta\varphi)^2+2K\Delta\varphi].\ee

As previously mentioned, $ a_{i}=\frac{\partial_{i} N}{N}$ is a vector field which arises due to non-projectability condition. Here $\Delta\equiv g_{ij}\nabla_{i}\nabla_{j}$ and $\Lambda_{g}$ is a coupling constant. 
The Ricci and Riemann tensors, $R_{ij}$ and $R^{i} _{jkl}$, are all made out of the $4$-metric $g_{ij}$. We also have
\be \Theta_{ij}=R_{ij}-\frac{1}{2}g_{ij}R+\Lambda_{g}g_{ij},\ee
and
\be K_{ij}=\frac{1}{2N}(-\dot{g}_{ij}+\nabla_{i}N_{j}+\nabla_{j}N_{i}),\ee
in the standard way. Also, $\hat\Theta^{ijkl}=g^{ik}g^{jl}-g^{ij}g^{kl}$.

Including all the relevant terms, the most general potential with the softly broken detailed balance condition is given by \cite{a.borzou,e.kiri}
\be\nonumber L_{V}=\gamma_{0}\zeta^2+(\gamma_{1}R-\beta_{0}a_{i}a^{i})+\frac{1}{\zeta^2}[\gamma_{2}R^2+\gamma_{3}R_{ij}R^{ij}\ee
\be\nonumber +\beta_{1}(a_{i}a^{i})^2+\beta_{2}(a^{i}_{\;i})^2+\beta_{3}(a_{i}a^{i})a^{j}_{\;j}\ee
\be\nonumber +\beta_{4}a^{ij}a_{ij}+\beta_{5}(a_{i}a^{i})R+\beta_{6}a_{i}a_{j}R^{ij}+\beta_{7}Ra^{i}_{\;i}]\ee
\be +\frac{1}{\zeta^4}[\gamma_{5}C_{ij}C^{ij}+\beta_{8}(\Delta a^{i})^2], \ee
where $C_{ij}$ is the five-dimensional Cotton tensor \cite{cotton} and $a_{ij}=\nabla_i \nabla_j \ln{N}$.
For future reference, we split up $L_V$ into two parts $L_V^R$ corresponding to $\gamma$-terms and $L_V^a$ corresponding to $\beta$-terms.

\section{The setup for brane solution}
\label{Sec2}

We are now able to look for braneworlds solutions in the aforementioned HMT setup with the non-projectability condition. Let us consider the usual Ansatz 
\be\label{metric-5d-geral} ds^{2} = e^{2A(w)}g_{\mu\nu}(t,x^{I})dx^{\mu} dx^{\nu}-dw^{2}\;\;\;(\mu,\nu=0,1,2,3),\ee
%The above equation can be written as
%\be ds^{2} = e^{2A(w)}(N^{2}(t,x^{I})dt^2+q_{IJ}(t,x^{I})dx^{I} dx^{J})-dw^{2},\ee
%or, equivalently
%\be ds^{2} = \bar{N}^{2}(t,x^{i})dt^2+\bar{q}_{IJ}(t,x^{i})dx^{I} dx^{J}-dw^{2},\ee
%where
%\be\bar{N}^{2}(t,x^{i})=\bar{N}^{2}(t,x^{I},w)=e^{2A(w)}N^{2}(t,x^{I}),\ee
%and
%\be \bar{q}_{IJ}(t,x^{i})=\bar{q}_{IJ}(t,x^{J},w)=e^{2A(w)}q_{IJ}(t,x^{I}).\ee
We are interested in studying a braneworld scenario in which the 3-brane is generated by a scalar field that depends only on the extra dimension $w$. That is, in the absence of such a field  this metric reduces to a five dimensional Minkowski spacetime. 

By using ADM decomposition in Eq.~(\ref{metric-5d-geral}) we simply identify $g_{00}=N^{2}=e^{2A(w)}$ and $g_{IJ}(t,x^{I})=diag(-1,-1,-1)$ (where $N_i=0$).
%In this sense we make $N(t,x^{I})=1$ ($\bar{N}^{2}=e^{2A(w)}$) and $g_{IJ}(t,x^{I})=diag(-1,-1,-1)$ ($\bar{q}_{ij}=e^{2A(w)}$). Thus, by considering all of this we obtain
%\be ds^2=e^{2A(w)}\eta_{\mu\nu}dx^{\mu}dx^{\nu}-dw^2,\ee
%with $\eta_{\mu\nu}=diag(1,-1,-1,-1)$. 
In this setup one can write the vector field $a_{i}=\partial_{i}{N}/{N}$ as
\be a_{i}=(0,0,0,A'(w)),\ee
where prime denotes derivative with respect to $w$. Moreover, with such metric the kinetic term turns out to be $L_{K}=0$ --- it is easy to check that Eq.~(\ref{Kinetic}) is identically null for static solutions. Furhtermore,  without loss of generality, one can choose the gauge $\varphi=0$ and $A=0$.
This is also in accord with equations of motion of the original Lagrangian. Notice that the equation of motion for $\varphi$ admit the solution $\varphi=0$ as one can be easily checked --- see \cite{apendix}. The equation of motion for $A$ is
\be\label{eq-A-1}
R-2\Lambda_g=8\pi G J_A, \qquad J_A=2\frac{\delta(NL_m)}{\delta A}, \qquad \zeta^2=\frac{1}{16\pi G }
\ee
which needs a nonzero current $J_A$ in order to allow for general spacetime solutions. This can be provided by adding such a current term in the matter sector as follows
\be\label{eq-A-2}
L_m\to L_m+\frac{A}{2N}J_A.
\ee
Thus, the solution $A=0$ is also satisfied by the equation of motions of the original Lagrangian --- see further below.

 Then, after these considerations, the action of the theory (\ref{theroy1}) takes the form
\be S=\zeta^2\int dtdx^{3}dw{N}\sqrt{g}\left(-L^{R}_{V}+\frac{1}{\zeta^2}L_{m}\right).\ee
The variation of this action with respect to ${N}$ gives rise to the following equation
\be\label{eqmov1}\zeta^2 (\sqrt{g}L^{R}_{V}+\sqrt{g}F_{V})=\frac{\delta S_{m}}{\delta {N}},\ee
where 
\be\label{lagRV} L^{R}_{V}=\gamma_{0}\zeta^2-R+\frac{1}{\zeta^2}(\gamma_{2}R^2+\gamma_{3}R_{ij}R^{ij})+\frac{\gamma_{5}}{\zeta^4}C_{ij}C^{ij},\ee
and $F_{V}$ is given by
\be\label{lagFV}\nonumber F_{V}=\beta_{0}(2a^{i}_{\;i}+a_{i}a^{i})-\frac{\beta_{1}}{\zeta^2}[3(a_{i}a^{i})^2+4\nabla_{i}(a_{k}a^{k}a^{i})]+\frac{\beta_{2}}{\zeta^2}\left[(a^{i}_{\;i})^2+\frac{2}{{N}}\nabla^2({N}a^{k}_{\;k})\right]\ee

\be\nonumber -\frac{\beta_{3}}{\zeta^2}\left[(a_{i}a^{i})a^{j}_{\;j}+2\nabla_{i}(a^{j}_{\;j}a^{i})-\frac{1}{{N}}\nabla^2({N}a_{i}a^{i})\right]+\frac{\beta_{4}}{\zeta^2}\left[a_{ij}a^{ij}+\frac{2}{{N}}\nabla_{j}\nabla_{i}({N}a^{ij})\right]\ee

\be\nonumber -\frac{\beta_{5}}{\zeta^2}[R(a_{i}a^{i})+2\nabla_{i}(Ra^{i})]-\frac{\beta_{6}}{\zeta^2}[a_{ij}a^{ij}R+2\nabla_{i}(a_{j}R^{ij})]+\frac{\beta_{7}}{\zeta^2}\left[Ra^{i}_{\;i}+\frac{1}{N}\nabla^2({N}R)\right]\ee

\be+\frac{\beta_{8}}{\zeta^4}\left[(\nabla a^{i})^2-\frac{2}{{N}}\nabla^{i}[\Delta({N}\Delta a_{i})]\right].\ee
Now taking the variation of the action of matter, we have 
\be \rho_{m}=-\frac{1}{\sqrt{g}}\frac{\delta S_{m}}{\delta {N}},\ee
where we define $\rho_{m}$ as the conventional matter and energy density. Thus, we can write (\ref{eqmov1}) as
\be L_{V}^{R}+F_{V}=-\frac{1}{\zeta^2}\rho_{m}.\ee
In the IR regime the forth and sixth spatial derivative terms can be neglected. This limit can be obtained by taking only the first two terms in (\ref{lagRV}) and the $\beta_{0}$ term in (\ref{lagFV}). 
Thus, making $\gamma_{0}=0$ we have %and $\gamma_{1}=-1$ we have the following action describing the IR behavior of the theory
\be R+\beta_{0}A'^2=\frac{1}{\zeta^2}\rho_{m},\ee
where the presence of $A'$ term reflects the nonprojectability condiction. Now, the variation of the action (\ref{theroy1}) with respect to $g_{ij}$ yields the dynamical equations
\be\label{eqmov2}\nonumber\frac{1}{{N}\sqrt{g}}\frac{\partial}{\partial t}(\sqrt{g}\pi^{ij})+2K^{ik}K^{j}_{k}-2\lambda KK^{ij}+\frac{1}{{N}}\nabla_{k}(\pi^{ik}N^{j}+\pi^{kj}N^{i}-\pi^{ij}N^{k})\ee
\be -\frac{1}{2}L_{K}g^{ij}-\frac{1}{2}L_{A}g^{ij}-F^{ij}-F_{a}^{ij}-F^{ij}_{\varphi}-\frac{1}{{N}}(AR^{ij}+g^{ij}\nabla^2A-\nabla^{j}\nabla^{i}A)=\frac{1}{2\zeta^2}\tau^{ij},\ee
where
\be \pi^{ij}=\frac{\delta({N}L_{K})}{\delta \dot{g}_{ij}}=-K^{ij}+Kg^{ij},\ee
\be\tau^{ij}=\frac{2}{\sqrt{g}}\frac{\delta(\sqrt{g}L_{m})}{\delta g_{ij}},\ee
\be F^{ij}=\frac{2}{\sqrt{g}}\frac{\delta(-\sqrt{g}L^{R}_{V})}{\delta g_{ij}}=\sum^{}_{s=0}\hat\gamma_{s}\zeta^{n_{s}}(F_{s})^{ij},\ee
\be F_{a}^{ij}=\frac{2}{\sqrt{g}}\frac{\delta(-\sqrt{g}L^{a}_{V})}{\delta g_{ij}}=\sum^{}_{s=0}\beta_{s}\zeta^{m_{s}}(F^{a}_{s})^{ij},\ee
and
\be F^{ij}_{\varphi}=\sum^{}_{s=0}\mu_{s}(F^{\varphi}_{s})^{ij}.\ee
%and
%\be F^{ij}_{A}=\frac{1}{\bar{N}}[AR^{ij}-(\nabla^{i}\nabla^{j}-g^{ij}\nabla^2)A].\ee
The expressions of $F_{s}$, $F_{s}^{a}$, and $F^{\varphi}_{s}$ can be found in Appendix of the reference \cite{apendix}. On the other hand, the coefficients are given by
\be\nonumber \hat{\gamma}_s=\left(\gamma_{0},\gamma_{1},\gamma_{2},\gamma_{3},\frac12\gamma_{5},-\frac{5}{2}\gamma_{5},3\gamma_{5},\frac{3}{8}\gamma_{5},\gamma_{5},\frac{1}{2}\gamma_{5}\right),\ee
\be\nonumber n_{s}=(2, 0, -2, -2, -4, -4, -4, -4, -4,-4),\ee
\be\nonumber m_{s}=(0, -2, -2, -2, -2, -2, -2, -2, -4),\ee
and
\be\mu_{s}=\left(2,1,1,2,\frac{4}{3},\frac{5}{3},\frac{2}{3},1-\lambda,2-2\lambda\right).\ee

Again, by the metric adopted and the gauge choice $A=\varphi=0$ one can write (\ref{eqmov2}) as
\be F^{ij}+F^{ij}_{a}=-\frac{1}{2\zeta^2}\tau^{ij}.\ee
Notice also that by using Eqs.~(\ref{eq-A-1}) and (\ref{eq-A-2}) the contribution of $L_A$ in (\ref{eqmov2}) is canceled out by an equivalent term in the matter sector.
In the IR limit, we have
\be\label{eqIR} \gamma_{0}\zeta^{2}(F_{0})^{ij}+\gamma_{1}(F_{1})^{ij}-\beta_0\frac{A'^2}{2}=-\frac{1}{2\zeta^2}\tau^{ij},\ee
where (see Appendix of \cite{apendix}, for example)
\be (F_{0})_{ij}=-\frac{1}{2}g_{ij},\ee
\be (F_{1})_{ij}=R_{ij}-\frac{1}{2}Rg_{ij}.\ee	
For $\gamma_{0}=0$ and $\gamma_{1}=-1$ (restriction due to observations in the IR regime of the theory) we have 
\be (F_{1})^{ij}+\beta_0\frac{A'^2}{2}=\frac{1}{2\zeta^2}\tau^{ij}.\ee
Finally, the ${44}-$component of this equation can be write as
\be R_{44}-\frac{1}{2}Rg_{44}+\beta_0\frac{A'^2}{2}=\frac{1}{2\zeta^2}\tau_{44}.\ee	

\section{Braneworlds from relativistic scalar field}
\label{Sec3}

We shall now consider a relativistic scalar field in the matter sector in order to find explicit braneworld solutions. 
The non-relativistic case is straightforward whose start point can be the general non-relativistic action given in \cite{e.kiri}.

\subsection{Field equations and first-order formalism}

\begin{figure}[h!]
		\includegraphics[scale=0.35]{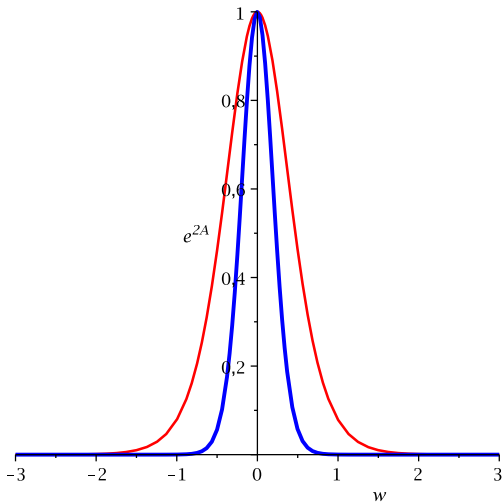}
		\caption{The warp factor $e^{2A(w)}$ (with $\lambda=1$) for $a = 3\sqrt{2}/4$ (red thin line) and $a = 3/2$ (blue thick line).}
	\label{fig1}
\end{figure}

Let us consider the following Lagrangian for the scalar field $\phi=\phi(y)$
\be L_{m}=L_{\phi}=\frac{1}{2}\epsilon g_{ab}\partial^{a}\phi\partial^{b}\phi-V(\phi),\ee
where $\epsilon=1$ for a standard dynamics and $\epsilon=-1$ for a ghost dynamics. 
We have $T_{00}=e^{2A}\left(\frac{1}{2}\epsilon\phi'^2+V(\phi)\right)$ and $T_{44}=\frac{1}{2}\epsilon\phi'^2+V(\phi)$ with $\rho_{m}=\rho_{\phi}=T^{0}_{0}$. 
Thus, the equations of motion read
\be\label{eq-45} -(12+\beta_{0})A'^2-6A''=\frac{1}{\zeta^2}\left(\frac{1}{2}\epsilon\phi'^2+V(\phi)\right),\ee
\be\label{eq-46} (6+\beta_{0})A'^2=\frac{1}{\zeta^2}\left(\frac{1}{2}\epsilon\phi'^2-V(\phi)\right),\ee
\be\label{eq-47} 6A''+6A'^2=-\frac{1}{\zeta^2}\epsilon\phi'^2.\ee
The scalar potential can be given by the general form
\be\label{pot-48} V=-\zeta^2(9 + \beta_0)A'^2 - 3\zeta^2 A''.\ee
Making $\beta_0=-6$  we find from (\ref{eq-46}) the following equation
\be\label{eq-49} V(\phi)=\frac{1}{2}\epsilon\phi'^2.\ee
In this sense we can identify the following first-order differential equations in terms of a general `superpotential' $W$ given by
\be\label{eq-50}\phi' = \frac18W_\phi,\ee
\be\label{eq-51} A' =-\frac13W,\ee
which solve Eqs.~(\ref{eq-45})-(\ref{eq-47}).
We may now consider the following simple and well-known example
\be\label{eq-52} V(\phi) = \frac12\lambda(a^2 - \phi^2)^2.\ee
Starting with equations (\ref{eq-49}) and (\ref{eq-52}), and making $\epsilon= 1$, we find
\be\label{eq-53} \phi = a \tanh{(a\lambda w)}.\ee
Now using (\ref{eq-49}) and (\ref{eq-50}) we have
\be\label{eq-54}  W_\phi =8a^2\lambda-8\lambda\phi^2, \ee
which allows us to find the superpotential 
\be\label{eq-55}  W(\phi) = 8a^2\lambda\phi - \frac{8\lambda}{3}\phi^3. \ee
Considering the equations (\ref{eq-51}), (\ref{eq-53}) and (\ref{eq-55}), we get the solution
\be A(w)= \frac{8a^2}{9}\ln[\tanh^2 {(a\lambda w)}-1]-\frac{4a^2}{9}\tanh^2{(a\lambda w)}.\ee
In Fig.~\ref{fig1} are depicted the behavior of the graviton wave function around the 3-brane which signalizes gravity localization. However, one should note from Eqs.~(\ref{eq-47})-(\ref{eq-49})
that for the kink solution (\ref{eq-53}), the potential (\ref{pot-48}) is asymptotically {\it flat}. This is completely differently of the Einstein gravity where this potential 
is expected to be asymptotically $AdS$ as well-known from Randall-Sundrum model in order for to have graviton zero mode localization.

\subsection{Dilatonic brane}

In order to find a solution to the equation (\ref{eq-47}) we use the Ansatz proposed in \cite{cs} --- see also \cite{brito-fonseca2,brito-fonseca,aqeel} for recent discussions --- 
in the context of the dilatonic domain wall
\be A(w)=B\ln{(1+cw)}, \qquad w>0,\ee
\be A(w)=B\ln{(1-cw)}, \qquad w<0. \ee
with $c>0$.
\begin{figure}[h!]
		\includegraphics[scale=0.35]{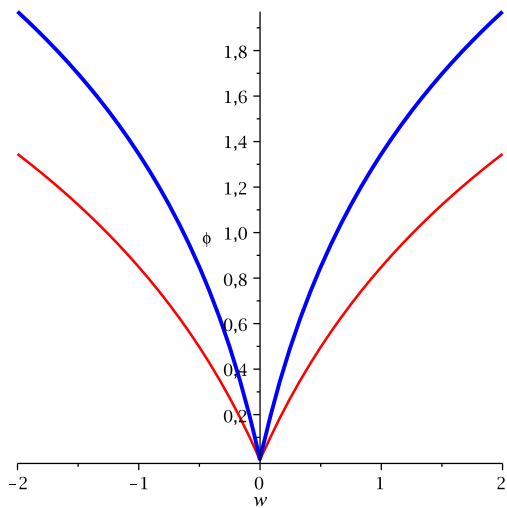}
		\caption{The solutions $\phi(w)$ (with $B = 1/2$, $\zeta=1$) for $c = 1$ (red thin line) and 2 (blue thick line).}
	\label{fig2}
\end{figure}
The `kink' profiles (with $0 < B < 1$) are given by
\be \phi(w)=\sqrt{6B(1-B)}\zeta\ln(1+cw), \qquad w>0\ee
and
\be \phi(w)=\sqrt{6B(1-B)}\zeta\ln(1-cw). \qquad w<0\ee
They are depicted in Fig.~\ref{fig2} for $c=1$ and $c=2$.
A scalar potential can be identified in this case as a usual dilatonic potential
\be V(\phi) = V_0e^{-\frac{2\phi}{\zeta\sqrt{6B(1-B)}}},\ee
where
\be V_0 = \zeta^2c^2B[3 - B(9 + \beta_0)].\ee
The Fig.~\ref{fig3} shows  the behavior of this potential for  $\beta_0 = -6$  and  $\beta_0 = -1$. It is easy to see that the potential is asymptotically flat and the warp factor diverges far from the brane. Thus, the localization of gravity can be achieved in this case only through metastable gravitons \cite{brito-fonseca2}.
\begin{figure}[h!]
		\includegraphics[scale=0.35]{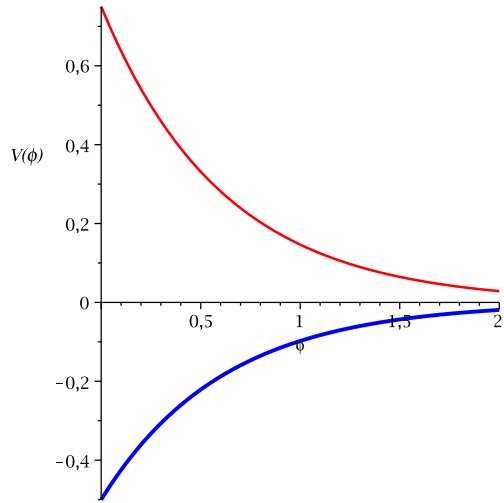}
		\caption{The dilatonic potential $V (\phi)$ (with $B=1/2,\zeta = c = 1$) for $\beta_0 = -6$ (red thin line) and $\beta_0 = -1$ (blue thick line).}
	\label{fig3}
\end{figure}

\subsection{Randall-Sundrum brane-like scenario}
In this model we simply have $A(w) = kw$ and $A'^2 = k^2\, (w\neq0)$ that integrating we find the well-known solution
\be A(w) = k |w| . \ee
For $w\neq 0$ we find $A''(w) = 0$. The potential (\ref{pot-48}) is now given by the simple form
\be V = -\zeta^2(9 + \beta_0)k^2. \ee
For $\beta_0 > -9$, $V$ is negative and plays the role of a five-dimensional cosmological constant in the bulk, which means an $AdS$ 
spacetime.

A real scalar field solution is possible to be found via Eq.~(\ref{eq-47}) if $\epsilon= -1$ (a `ghost' dynamics), for $\zeta$ being a real constant, such that 
\be \phi(w) = \sqrt{6}k\zeta w.\ee

\section{Conclusions}
\label{conclu}

In this paper we have found braneworld solutions in an alternative theory of gravity in the realm of HMT with non-projectable condition. A relativistic scalar
field is considered in the matter sector, and for some specific set of parameters we have been able to find
solutions of the equations of motion that solve first-order differential equations. The bulk is mostly asymptotically flat except in the case of the Randall-Sundrum brane-like scenario, whose related scalar field admits a `ghost' dynamics. One may extend our analysis to several models with one or more scalar fields.

\acknowledgements

We would like to thank to CNPq, PNPD-CAPES, PROCAD-NF/2009-CAPES for partial financial support.


\begin{thebibliography}{99}

\bibitem{rand}L. Randall and R. Sundrum, Phys. Rev. Lett. {\bf83}, 4690 (1999).
\bibitem{gold-wise} W. D. Goldberger and M. B. Wise, Phys. Rev. Lett. {\bf83}, 4922 (1999).
\bibitem{cs} M. Cvetic and H. H. Soleng, Phys. Rev. D {\bf51}, 5768 (1995);
Phys. Rep. {\bf282}, 159 (1997); 
\bibitem{varios1} K. Skenderis and P. K.
Townsend, Phys. Lett. B {\bf468}, 46 (1999); O. DeWolfe,
D. Z. Freedman, S. S. Gubser, and A. Karch, Phys. Rev. D
{\bf62}, 046008 (2000); C. Csaki, J. Erlich, T. Hollowood, and
Y. Shirman, Nucl. Phys. B {\bf581}, 309 (2000); C. Csaki, J.
Erlich, C. Grojean, and T. Hollowood, Nucl. Phys. B {\bf584},
359 (2000); M. Gremm, Phys. Lett. B {\bf478}, 434 (2000); M.
Porrati, Phys. Lett. B {\bf498}, 92 (2001); J. F. Vazquez-Poritz,
J. High Energy Phys. {\bf0112}, 030 (2001); {\bf0209}, 001 (2002); F.A.
Brito, M. Cvetic, and S.-C. Yoon, Phys. Rev. D {\bf64}, 064021
(2001); M. Cvetic and N. D. Lambert, Phys. Lett. B {\bf540},
301 (2002); A. Melfo, N. Pantoja, and A. Skirzewski,
Phys. Rev. D {\bf67}, 105003 (2003); D. Bazeia, F. A. Brito,
and J. R. Nascimento, Phys. Rev. D {\bf68}, 085007 (2003); D.
Bazeia, C. Furtado, and A. R. Gomes, J. Cosmol.
Astropart. Phys. {\bf0402}, 002 (2004); D. Bazeia and A. R.
Gomes, J. High Energy Phys. {\bf0405}, 012 (2004); O.
Castillo-Felisola, A. Melfo, N. Pantoja, and A. Ramirez,
Phys. Rev. D 70, 104029 (2004); K. Takahashi and T.
Shiromizu, Phys. Rev. D {\bf70}, 103507 (2004); R. Guerrero,
R. Omar Rodrigues, and R. Torrealba, Phys. Rev. D {\bf72},
124012 (2005).

\bibitem{varios2} V. Dzhunushaliev, Gravitation Cosmol. {\bf13}, 302 (2007); A.
de Souza Dutra, A. C. Amaro de Faria, Jr., and M. Hott,
Phys. Rev. D {\bf78}, 043526 (2008); Y.-X. Liu, L.-D. Zhang,
L.-J. Zhang, and Y.-S. Duan, Phys. Rev. D {\bf78}, 065025
(2008); C.A. S. Almeida, M. M. Ferreira, Jr., A. R. Gomes,
and R. Casana, Phys. Rev. D {\bf79}, 125022 (2009); Y.-X. Liu,
Z.-Hua Zhao, S.-W. Wei, and Y.-S. Duan, J. Cosmol.
Astropart. Phys. {\bf0902}, 003 (2009); Y.-X. Liu, H.-T. Li,
Z.-H. Zhao, J.-X. Li, and J.-R. Ren, J. High Energy
Phys. {\bf0910}, 091 (2009); A. E.R. Chumbes, A. E. O.
Vasquez, and M. B. Hott, Phys. Rev. D {\bf83}, 105010
(2011); A.C. Correa, A. de Souza Dutra, and M. B.
Hott, Classical Quantum Gravity 28, 155012 (2011);
  D.~Bazeia, F.~A.~Brito and F.~G.~Costa,
  Phys.\ Rev.\ D {\bf 87}, 065007 (2013).

\bibitem{varios3} D. Z. Freedman, C. Nunez, M. Schnabl, and K. Skenderis,
Phys. Rev. D {\bf69}, 104027 (2004); A. Celi, A. Ceresole, G.
Dall'Agata, A. Van Proeyen, and M. Zagermann, Phys.
Rev. D {\bf71}, 045009 (2005); M. Zagermann, Phys. Rev. D
{\bf71}, 125007 (2005); D. Bazeia, C. B. Gomes, L. Losano,
and R. Menezes, Phys. Lett. B {\bf633}, 415 (2006); V.I.
Afonso, D. Bazeia, and L. Losano, Phys. Lett. B {\bf634},
526 (2006); K. Skenderis and P. K. Townsend, Phys. Rev.
Lett. {\bf96}, 191301 (2006); D. Bazeia, F. A. Brito, and L.
Losano, J. High Energy Phys. {\bf0611}, 064 (2006); K. Skenderis
and P. K. Townsend, J. Phys. A {\bf40}, 6733 (2007).

\bibitem{varios4} A. Ceresole and G. Dall'Agata, J. High Energy Phys. {\bf0703},
110 (2007); W. Chemissany, A. Ploegh, and T. Van Riet,
Classical Quantum Gravity {\bf24}, 4679 (2007); E.A.
Bergshoeff, J. Hartong, A. Ploegh, J. Rosseel, and D.
Van den Bleeken, J. High Energy Phys. {\bf0707}, 067 (2007);
L. Cardoso, A. Ceresole, G. Dall'Agata, J. M. Oberreuter,
and J. Perz, J. High Energy Phys. {\bf0710}, 063 (2007); B.
Janssen, P. Smyth, T. Van Riet, and B. Vercnocke, J.
High Energy Phys. {\bf0804}, 007 (2008); M. Cvetic and M.
Robnik, Phys. Rev. D {\bf77}, 124003 (2008); S. Ferrara, A.
Gnecchi, and A. Marrani, Phys. Rev. D {\bf78}, 065003 (2008);
Y.-X. Liu, L.-D. Zhang, L.-J. Zhang, and Y.-S. Duan,
Phys. Rev. D {\bf78}, 065025 (2008).

\bibitem{horava}P. Horava, Phys. Rev. D {\bf79}, 084008 (2009).
\bibitem{blas2}D. Blas, O. Pujolas and S. Sibiryakov, JHEP {\bf1104}, 018 (2011); A. Padilla, J. Phys. Conf. Ser. {\bf259}, 012033 (2010); T.P. Sotiriou, J. Phys. Conf. Ser. {\bf283}, 012034 (2011); 
M. Visser, J. Phys. Conf. Ser. {\bf314}, 012002 (2011); P. Horava, Class. Quant. Grav. {\bf28}, 114012 (2011); T. Clifton, P.G. Ferreira, A. Padilla and C. Skordis, Phys. Rept. {\bf513}, 1 (2012).
\bibitem{b.pere}B. Pereira-Dias, C. Hernaski and J. Helayel-Neto, JHEP {\bf1203}, 013 (2012).
\bibitem{s.muk}S. Mukohyama, Class. Quant. Grav. {\bf27}, 223101 (2010).
\bibitem{cosmovarios} A. Wang and Q. Wu, Phys. Rev. D 83 (2011) 044025. K. Izumi and S. Mukohyama, 
Phys. Rev. D {\bf84}, 064025 (2011); A.E. Gumrukcuoglu, S. Mukohyama and A. Wang, Phys. Rev. D {\bf85}, 064042 (2012); D. Salopek and J. Bond, Phys. Rev. D {\bf42}, 3936 (1990); D.H. Lyth, K.A. Malik and M. Sasaki, JCAP {\bf0505}, 004 (2005).
\bibitem{hmt} P. Horava and C.M. Melby-Thompson, Phys. Rev. D {\bf82}, 064027 (2010).
\bibitem{blas} D. Blas, O. Pujolas, and S. Sibiryakov, Phys. Rev. Lett. {\bf104}, 181302 (2010); JHEP {\bf1104}, 018 (2011).
\bibitem{kimpton} I. Kimpton and A. Padilla, JHEP {\bf1007}, 014 (2010).
\bibitem{m.li} M. Li and Y. Pang, J. High Energy Phys. {0908}, 015 (2009); M. Henneaux, A. Kleinschmidt, and G.L. Gomez, Phys. Rev. D {\bf81}, 064002 (2010).
\bibitem{j.kluson} J. Kluson, JHEP {\bf1007}, 038 (2010).
\bibitem{t.zhu} T. Zhu, Q. Wu, A. Wang, and F.-W. Shu, Phys. Rev. D {\bf84}, 101502 (R) (2011).
\bibitem{a.borzou} A. Borzou, K. Lin, and A. Wang, JCAP {\bf1105}, 006 (2011).
\bibitem{apendix} T. Zhu, F.-W. Shu, Q. Wu and A. Wang, Phys. Rev. D {\bf85}, 044053 (2012).
\bibitem{e.kiri} E. Kiritsis and G. Kofinas, Nucl. Phys. B {\bf821}, 467 (2009).
\bibitem{benfica}F.S. Bemfica, M. Dias, M. Gomes, J.M. Hoff da Silva, EPJC {\bf73}, 2376 (2013). 

%\cite{Garcia:2003bw}
\bibitem{cotton} 
  A.~Garcia, F.~W.~Hehl, C.~Heinicke and A.~Macias,
  %``The Cotton tensor in Riemannian space-times,''
  Class.\ Quant.\ Grav.\  {\bf 21}, 1099 (2004)
  [gr-qc/0309008].
  %%CITATION = GR-QC/0309008;%%
  %50 citations counted in INSPIRE as of 31 Dec 2014

\bibitem{brito-fonseca2}R.~C.~Fonseca, F.~A.~Brito and L.~Losano, JCAP {\bf 1201}, 032 (2012).
\bibitem{brito-fonseca}R.~C.~Fonseca, F.~A.~Brito and L.~Losano, Phys.\ Lett.\ B {\bf 728}, 443 (2014).
\bibitem{aqeel}A. Ahmed, B. Grzadkowski and J. Wudka, JHEP {\bf1404}, 061 (2014).
\end{thebibliography}
\end{document}